# Universal Service: specific services on generic networks – some logic begins to emerge in the policy area: Richard A. Cawley

**Abstract**

*There is an extensive literature on many aspects of universal service, much of it dealing with aspects of policy [examples: historical lessons and network development (Mueller, Gabel), efficiency and pricing (Katz, Crandall, Waverman, Hausman, Riordan), models for universal service including financing issues (Schement, Noam, Weller), cost modelling issues (Hatfield, Gabel, Sharkey, FCC), regulatory problems and relationship with interconnection (Wildman, Weinberg), empirical surveys (Greenstein, NTIA), assignment, auctions, network effects (Tirole, Cremer)].*

*A major problem has been to translate the lessons from the literature into the policy and regulatory framework because of political interests and regulatory capture. Neither the USA or Europe has made a very good job of devising a 'clean' framework and the WTO agreement is sparing in this area. However, a number of pressures in the European context have enabled a more systematic approach to emerge, which also exploits the academic work. The pressures include (a) the need for the European policy and regulatory framework on universal service to encompass E. European countries where network development and income levels are much lower (b) the desire to encompass Internet in some form within the universal service regulatory framework (c) a willingness to design a regulatory framework that covers all communications networks and remove the telecommunications bias, thereby forcing issues of economic neutrality to the fore as well as adding other difficult public interest issues from the broadcasting area.*

*The increasing complexity of practically implementing costing and financing mechanisms (as well as the complexity of actually defining the elements covered by universal service obligations) has brought to the fore issues raised before in the literature with respect to various aspects of neutrality and economic efficiency. The paper systematically goes through a number of key areas and principles of regulation and how they are being designed to deal with a range of national situations. They include: defining the scope of universal service and the principles by which it might be modified; incorporating latitude for intervention outside this defined scope, defining incentive and designation methods to encourage the efficient supply of elements of universal service obligations; interpreting affordability in the context where price and income levels can diverge considerably; designing efficient costing methods where they are necessary; formulating alternative financing methods including central government financing and value added tax type methods which can co-exist and provide comparative*

*policy yardsticks; and finally designing appropriate 'must carry' policies for networks so as to incorporate public interest issues stemming from the broadcasting side.*

**Universal Service: specific services on generic networks – some logic begins to emerge in the policy area: Richard A. Cawley**

**Introduction and Aim**

Despite the wealth of literature (theoretical, descriptive and empirical) concerning universal service, the great challenge for policy makers has been to translate theory into practice in this area. It is very difficult to dismantle inefficient systematic price and other distortions in the face of political opposition. The price distortions themselves act as a barrier to entry and prevent the kind of competition and innovation that can help to address some of the problems. Very often, an historical social and/or regulatory contract has been established that guarantees universal provision in exchange for a privileged supply position. It is difficult to induce monopoly or dominant suppliers to offer choice or price discriminate in a way that confers benefits to more vulnerable consumers. Where universal service guarantees or safety nets generate net costs, it is also difficult to install financing mechanisms that are economically neutral and efficient such as via government budgets.

The purpose of this paper is to document the legal and policy changes underway in Europe and in particular to examine a number of specific issues that are being tackled to (i) put universal service provisions on a sounder economic policy basis (ii) set universal service in the context of policy for a broader communications sector (iii) establish flexibility (and its limits) for dealing with universal service guarantees in economies with significant variations in levels of telecommunications usage and income, as will be the case in an expanded European Union.

**General Background to Universal Service Legislation in the European Context**

European legislation on universal service defines, at a European Union (EU) level, "the minimum set of services of specified quality available to all users and consumers regardless of their geographic location, in the light of specific national conditions, at an affordable price, without distorting competition". These set of services and the obligations on specified or designated undertakings or operators that go with them are termed universal service obligations (USOs).

There are four specific aspects to the universal service regulatory framework at EU level.
- The specific definition of the minimum set of services

- Specification of quality (e.g. things like guarantees on delivery time of a connection, response time for directory enquiries, working order of payphones etc)
- Guarantees on price and affordability. Affordability has two aspects, the level of prices or expenditure and user control.
- A process for designating USO operators and a process for compensation of any net costs which may arise due to the minimum set of services being provided to (some) users at prices which are below commercial levels.

Certain aspects of the framework must be uniformly and commonly applied. These include the definition of the minimum scope of services, the specification of quality (of universal services provided by the designated operators), the measures designed to empower users to monitor and control expenditure, and the approach and method used (if necessary) to calculate any net costs of USOs. The process for designating undertakings required to provide some or all of the elements of universal service is subject to a number of criteria and requirements. These are dealt with in more detail below.

Member states have flexibility in two key areas. Subject to general requirements for tariff re-balancing and avoidance of anti-competitive tariff distortions, they oversee the level and structure of tariffs (including low user tariffs or variable two-part (fixed and variable) tariff options) that are appropriate in the national context to deal with affordability. Secondly, they can determine, within certain constraints, the means of recovering any net costs.

**Market Distortions and Competitive Distortions**

The first general distinction made in the new European universal service legislation is between market distortions and distortions of competition. The requirement is to minimise the first and avoid the second. This may seem an overly subtle point. However, it is important in the following sense. Intervention in the form of requiring a designated universal service provider to set common or average tariffs or providing service below cost for certain end users is likely to involve some form of market distortion whereby goods are supplied to final consumers under conditions that are different from that which emerges under market forces. Such distortions may be minimised, for example, by only subsidising marginal users and not infra-marginal ones. At the same time, any regulated requirement that involves subsidisation is likely to lead to a net cost incurred by the designated provider. In order to avoid competitive distortions between providers, undertakings incurring net costs need to be compensated. Hence the two general principles that are mutually consistent: minimum market distortion and no distortion of competition.

These principles are also stronger than those established in the commitment made in the context of the WTO agreement applying to the telecommunications sector. Here the reference in the context of members defining and maintaining universal service obligations refers to "competitively neutral manner" and "not more burdensome than necessary for the kind of universal service defined".

**Universal service scope – technological neutrality**

The existing European legislation refers to the provision of a voice telephony service (defined in the law) by fixed telecommunications operators. The new framework attempts to establish a technological neutral approach by dropping this definition and simply referring to the provision of access and service at a fixed location or address. This enables traditional wire-line technologies to be used to provide the defined universal service but also clearly permits other wire-line and wireless technologies to be used.

The exploitation of cellular wireless technologies to provide universal service raises some interesting issues. Already in Spain, Telefonica's mobile cellular network, albeit in a modified form, is used to deliver traditional 'fixed' telephone service. The subscribing household is furnished with a traditional termination point, to which it can connect standard telephone handsets (corded or cordless). The termination point links to a local exchange or concentration point via cellular wireless. No roaming capability (beyond the home cell) is provided. The provision of service at fixed locations via such methods, or indeed via other wireless technologies is also developing in central and E. European countries. The first issue concerns the use of the connection for data services that are included in the defined scope of universal service. This issue is treated in the next section.

The second issue concerns whether the provision of service via cellular wireless or mobile technology, without the provision of the 'fixed' termination point, constitutes a fulfilment of the defined universal service scope. The tariff or set of prices provided as part of the service could be structured in a way that fulfils the affordability criteria in the universal service definition. However, an intrinsic part of universal service involves having a ('fixed') telephone number that is part of the local numbering area. This implies a given structure of tariffs for both outgoing and incoming calls. This issue has not been explicitly covered in the new legislation. However, member states have flexibility here to fully exploit the possibilities for cellular wireless technologies and services provided by mobile operators to provide an alternative means (to the two methods above) of offering the defined universal service at fixed locations or addresses.

**Universal service – flexibility within the defined scope**

The defined minimum scope of universal service in current legislation includes data service (and implicitly dial-up Internet access) as well as telephone service. The new framework explicitly refers to Internet (in fact narrow-band access) but without specification of a given data rate. There is explicit flexibility up to narrow-band capacity or a data rate of 56kbit/sec. This is designed to deal with two practical problems.

The first is the case (as with Spain) of providing telephone service over modified mobile cellular networks. The key issue here is that data rates in general are lower than those experienced over copper loops. Therefore flexibility is given to permit the upgrading of networks to provide equivalent data rates or data rates up to the maximum narrow-band rate. In circumstances where such network upgrading leads to a net cost, this may be recovered by the financing methods permitted for universal service costs.

The second issue concerns, in particular, the countries of central and eastern Europe (C and EE countries) who will in the future join the European Union, but it also raises a broader issue of regulatory incentives and remedies. Telephone penetration in C and EE countries is still low compared to western Europe. However, coverage of mobile networks in many countries exceeds 95% of the population and take-up of mobile subscriptions is growing fast and often exceeds traditional fixed line use. Setting a rigid requirement for the functionality of Internet access within the defined universal service requirement would prevent mobile services from fulfilling the universal service criteria and would remove any incentives for undertakings investing in mobile networks to compete to provide services for users covered by or potentially covered by the universal service guarantees. In the future, depending on the deployment of upgrades to second generation mobile technology or on investments in third generation, cellular wireless networks may be able to offer comparable data rates.

In order to preserve the incentives of non incumbent undertakings and undertakings using other technologies to compete to provide the guaranteed services within universal service, specific flexibility on data rates is built into the newly proposed legislation.

**Universal service – flexibility outside the defined scope**

The current legislation recognises that member states may wish or need to take measures outside the common minimum guarantee of universal service that exists at an EU wide level. Such intervention could address a range of policy concerns including provisions directed at public use (schools, libraries) that go beyond private use, special social measures for equipment or service that go beyond those covered in the EU legislation or specific regional measures. The newly proposed framework further clarifies and refines what is possible outside the commonly defined scope.

Member states may decide to guarantee the provision of additional services, beyond those covered by the universal service obligations but in such cases no compensation from within the sector may be imposed. Any financing requirement must come from the government budget. In addition any additional measures (e.g., facilitating the development of infrastructure or services in circumstances where the market does not satisfactorily address needs), must be in conformity with Community law. This means that such measures must be notified as state aid measures under the Treaty. They can then be assessed for approval or derogation under the Treaty in a number of ways. Such derogation may be under article 86 (services of general economic interest) or under other competition articles of the Treaty including state aid where the case law has established a number of categories and criteria for permitting public intervention or subsidy.

**Designation of universal service undertakings and incentives**

Some of the economic literature on universal service extols the virtues of using auctions to assign USOs to designated undertakings. There has also been extensive work on appropriate auction design in this area. But it is also evident that considerable problems may arise. This is due in part to the difficulty of ensuring that sufficient undertakings are in a position to bid against the incumbent (they would need to use alternative network technologies or acquire the use of the assets of the incumbent) and because of the asymmetries of information (for example concerning the net costs or benefits of serving groups of subscribers) between the incumbent and potential entrants.

The proposed legislation is therefore designed to cover a range of approaches to designating USO undertakings from situations whereby the incumbent automatically acquires the right and obligation, through to tendering methods and auctions. The variations that are possible also affect related measures to assess any net cost of providing universal service. If the obligation is acquired by tendering or auction, the net cost is determined at the same time. If the obligation is assigned by another means, there may or may not be a need to undertake a calculation of net cost. An intermediate possibility exists where the obligation is assigned to the incumbent with the regulatory condition that the undertaking itself must if it wishes request a compensation. Such a request would trigger a process by which the regulator decides whether to assign the obligation to another undertaking(s) or alternatively undertake an assessment of the net cost borne by the incumbent.

The legislation makes clear that no undertaking is a priori excluded from being designated. This avoids the necessity of automatically launching an open process for designation, where it is unnecessary or cumbersome. At the same time it maintains pressure on the incumbent. In particular the possibility that the separate elements of the USO may be assigned to different undertakings or that the geographic coverage of the obligation may be sub-divided also encourages efficiency. The various possibilities and stages in a designation process and their link to costing and financing stages are illustrated in annex 1.

**Competitive neutrality**

There are a number of aspects of neutrality that need to be considered when designing universal service policies, as has been pointed out in some of the literature. The section above already made the distinction between minimising distortions on final consumption and avoiding distortions between undertakings. Because USOs may lead to net costs that require financing, it is important to try and minimise losses of allocative efficiency due to undertakings passing on this burden to consumers via price increases. One way to achieve this is to finance from the government budget. An alternative is to recover the amount via a neutral tax over the broadest possible base. These financing issues are discussed in the following section.

The other key aim is to avoid or at least in practice to minimise distortions to the competitive process and avoid discrimination. Any contributions that need to be raised to compensate for net costs should be neutral with regard to market players, to services, to vertical structures and to technology. Neutrality with regard to market players implies that financing should be directly related to economic activity. Neutrality with regard to services means the avoidance of favouring particular services or activities over others. Neutrality with regard to vertical structure requires the avoidance of accumulated contributions; e.g. a service provider paying on the basis of its own activities and in relation to inputs that it purchases from other operators. Technological neutrality means that contributions should not for example distinguish between alternative transmission modes.

As indicated above, financing any net costs via central budgets avoids many of these neutrality problems arising (unless there are significant inefficiencies with current taxation and expenditure structures). However, given political considerations it is often necessary to compensate net costs borne by a designated USO undertaking from within the sector.

A key issue has arisen in the European context as to whether competitive distortions may arise when one country finances a net USO cost from central government budgets and another finances from within the sector. The former arrangement results in the undertaking receiving from the government budget an amount equivalent to the compensation amount that has been determined (via an auction or tendering process or via a regulatory net cost calculation). This is tantamount to treating the USO obligation and its potential compensation as a public procurement decision by the government.

In the second case, the undertaking is compensated from a virtual or explicit sector fund to which all eligible undertakings contribute on the basis of a contribution rule (e.g. based on net turnover or value added). In essence the USO undertaking receives from other undertakings in the sector a part of the net cost burden in proportion to its own economic activity. If it has half of the net turnover with respect to the eligible base for contributions it receives half the compensation amount from other undertakings. Might this difference in compensation approach lead to competitive distortions between countries (something that the Treaty of the European Union, in particular the articles concerning competition and internal market policy, expressly forbids).

The first point to make is that the first potential source of economic or competitive distortion is likely to come from the institution in the first place of a universal service compensation arrangement and in particular, one where the net cost is calculated. There is always a high risk that the determined or calculated cost will be much higher than the actual cost. Therefore the choice of financing via government budget rather than within the sector seems more likely to add incentives to discipline the process. Such discipline may be weaker in the sector compensation approach unless the incumbents' competitors are strong relative to the incumbent.

The second point is that within a given country, the two approaches (sector compensation or finance from the government budget) can be designed so as to give equivalent effects. This would be so if in the first case, the amount is recovered via an incremental tax on value added within the sector and the required amount is simply deducted from the full amount of value added tax that is normally transferred from the sector to government receipts. This is shown in more detail in annex 2.

The third point is that in principle the two different approaches to compensation could induce different financial effects on the undertakings concerned. But the same problem may occur in principle in comparing a country with a USO cost burden and one without. Such differences are therefore akin to divergences in other economic factors such as labour costs, costs of capital (borrowing and equity) and taxation regimes. Fortunately, universal service net cost burdens are small relative to overall economic activity and even to turnover of the undertakings concerned. Moreover, undertakings in the sector compete in general within national markets even if communications is by definition an international activity. Therefore the key concern should be to ensure that designated USO undertakings are adequately compensated (but no more than adequately compensated) for the net cost burden on any given national market. It is this key principle that is part of the legislation although the need to assess and avoid competitive distortions between countries has also been recognised.

**Financing or cost recovery**

Existing legislation envisages that any net costs produced by USOs will be financed from within the sector. Two general approaches are possible, one involving recovery or levies on companies or undertakings, the other involving levies on end users directly. Only the former has been used to date. In the former case, undertakings will in turn decide how they generate or recover the contributions. Presumably, this will involve recovery in economic areas where they face the least competition or from services where their customers have inelastic demand.

Certain principles have been written into the legislation or expounded in subsequent recommendations. The initial proposals of the Commission with respect to legislation in the new framework attempted to reinforce these principles of neutrality and efficiency. One key proposal was that clear preference be given to funding any determined net USO costs from government budgets. This proposed preference failed and the newly proposed legislation retains the two options for financing as indicated above.

A second proposal was to use a type of value added tax mechanism for financing any net USO cost where this was undertaken from within the sector. The sensitivity of member states to any proposal connected to existing taxation systems or methods (an area where they individually have the right of veto over legislative changes) meant that this proposal failed. The Commission has however, pointed out the possible difficulty in the future, (as services and networks continue their convergence, more diverse companies enter the sector and vertical linkages become more rather than less complex), of recovering net USO costs on the basis of levies on companies. A value added tax type mechanism is likely to be more practical and transparent as well as more efficient and neutral in this respect.

Despite the lack of success in establishing a value added type tax mechanism where USO cost is recovered within the sector, it has been possible to establish the importance of minimising the impact on consumption decisions. In the current legislation, cost recovery is based on a link between the services subject to universal service provisions (essentially voice telephony as defined in the legislation) and the undertakings providing such services - public telecommunications network operators and voice telephony service providers. Several changes appear in the newly proposed legislation. First the notion of voice telephony as a defined service disappears. Secondly with respect to sharing any net USO cost within the sector, the reference is to a broader category of undertakings - providers of electronic communication networks and services. Thirdly sharing mechanisms must respect principles of transparency, least market distortion, non-discrimination and least market distortion. Least market distortion is to be interpreted as recovery in a way that as far as possible minimises the impact of the financial burden falling on end-users, for example by spreading contributions as widely as possible.

**Universal service – reviewing the defined scope**

Prior to the policy review that led to the new legislative proposals for the European regulatory framework, pressure had been building up to review and indeed extend the minimum universal service guarantee. The initial pressure concerned possible inclusion of Internet access and services for schools (beyond that already implicitly included in the scope). More recently the possible extension to broadband Internet has been raised.

It has been argued strongly why it was premature to extend the defined scope to any form of broadband Internet access. Because the European legislation also guarantees affordability, this would raise the prospect of a substantial increase in net USO cost. For example a 10 euro subsidy per month of the next 10% increment of households subscribing to broadband Internet would generate a potential net cost of over 2 billion euros. If this burden were to be recovered from the whole sector, average and lower income households would be subsidising the early takers, irrespective of the distortions arising from subsidising particular undertakings or particular ways of providing broadband Internet.

The solution then to the pressure to extend universal service has been to include a review process in the legislation. The review is to be undertaken periodically (every three years) in the light of economic, social and technological developments. Any possible modification to the guaranteed scope of universal service at EU level is subject to the following twin test:

- are specific services available to and used by a majority of consumers and does the lack of availability or non-use by a minority of consumers result in social exclusion, and
- does the availability and use of specific services convey a general net benefit to all consumers such that public intervention is warranted in circumstances where the specific services are not provided to the public under normal commercial circumstances?

In essence, therefore, the review criterion is a combination of a majority use test and a market failure one.

**Availability and Affordability**

One area that has always been a source of contention and confusion is the issue of affordability. In European legislation, affordability is the key part of the minimum universal service guarantee. Therefore universal service is not simply about guaranteeing availability (typically the case with public broadcasting where incremental costs are often very low) but includes the twin guarantee of affordability.

Affordability tends to be more complex in the provision of telecommunications services where two part tariffs are common. In theory, any tariff with a non zero connection or fixed charge could discourage a user joining a network and thereby justify a subsidy in order that social and private benefits converge. In practice, empirical evidence shows that many are disconnected or are discouraged from joining a network because of usage charges or the inability to control them.

Some have called for affordability to be determined at a European Union level. This has been resisted, for obvious reasons, bearing in mind that problems at the margin (in any national context) can be dealt with in a range of ways, for example tariff options, special low user schemes, vouchers etc.

The issue of affordability has arisen again more recently in connection with plans for European Union enlargement to include countries where income levels are generally much lower than in the existing 15 member states and where fixed telephone penetration is also much lower. Incidentally such historically low levels (even with respect to income levels) has often been the result of deliberate regulatory intervention either to curtail access or to set prices at unrealistic rates such that waiting lists have built up and incumbent operators have had little incentive to invest and serve users.

It has been argued that the affordability requirement in the minimum EU universal service guarantee implies that heavy subsidies are needed so that all users not currently subscribing are connected. However, documented developments in other European countries show that some users are substituting mobile for fixed access when subscribing to telephone service (e.g. in 1999 in Portugal, of the 81% of households with telephone service, 12% of total households only have a mobile subscription, and a further 37% have both). In addition, depending on the roll-out of digital terrestrial and satellite TV services or the upgrading of cable TV networks, significant proportions of households are getting email or Internet access via non traditional telecommunications networks. Therefore, it is important to maintain incentives for competing networks or technologies to provide part or the entire universal service guarantee.

Secondly, there is no obvious reason to subsidise or maintain artificially distorted tariff structures in the belief that this is the only way to increase the user population, in circumstances in which rates of user growth continue to remain at or above historic growth rates. In that sense affordability should be interpreted as affordability at the margin or at least amongst a significant incremental group of users who are considering taking up or dropping service. It should not be understood in the sense of affordability for example for the 50% of the population who have not yet realistically considered taking up service.

A more general political and economic question nevertheless arises in the context of affordability and availability. When legislation on universal service is passed or adopted, newspapers do not normally distinguish between availability and affordability guarantees. It is very difficult to argue that the fact that a service is not included in the universal service guarantee (because the affordability guarantee then means that a subsidy or financing mechanism is required), does not mean that it will not be widely available to users if they care to pay for it. In other words, there is a major difference between guaranteed nation-wide availability of broadband Internet access (albeit at commercial prices) and a universal service guarantee.

The only realistic way to deal with such a problem of political economy is to distinguish clearly and legally between universal availability and the minimum universal service guarantee itself, the latter being explicitly linked to possible costing and financing requirements. This distinction is already implicit in the European legal measures requiring the retail provision of a minimum set of leased lines. Given the political pressures surrounding the development of Internet access, it seems inevitable that such a distinction will need to be made in the legislation in the next few years.

**Conclusions**

The aim of this paper has been to document (in a fairly simple and superficial way) the progress in injecting some economic logic and lessons from the historical literature into the policy process for universal service in the European context. Despite the political pressures and despite the desire in certain countries to continue to undertake distribution and social policy via sector intervention, experience shows that it is possible to construct legal and regulatory frameworks that properly and rightly address social concerns and the interests of users that are vulnerable to market forces, whilst maintaining economic and regulatory incentives and addressing efficiency issues. It would have been inconceivable a few years ago to envisage legislation that specifically encompassed the financing of universal service net costs from government budgets. Nor that the issue of allocative efficiency is implictly treated. The fact that this is now the case shows that much progress has been made.

# Annex 1: Designation and Costing/Financing Process

**Designation of USO undertakings**

- in considering designation, separate USO elements (incl. geographical coverage).
- may not need to designate and/or determine whether any net cost [e.g. incentive process, request by historical USO provider].

**Designation – YES**

- designate but determine net cost separately.
- designate and determine net cost at same time (via aucton or tender).

**Net Costing Determination**

- not needed.
- NRA calculation required.
- determined as part of designation process.

**Net Cost Calculation Required**

- calculation may lead to cost that is low relative to administrative cost of setting up financing arrangement.
- net cost calculated and proceed to financing arrangement.

**Financing or Recovery of Net Cost**

- may not be needed.
- recovery via government funding only.
- recovery via sector only.
- combination of above.

**Recovery from within sector**

- recovery from undertakings directly.
- recovery from end-users directly.
- For both cases, the base is relevant, as is neutrality.

**Recovery form undertakings directly**

- many possibilities but few are neutral - gross revenues, profits, traffic all lead to distortions
- exempt companies below given level of economic activity.
- best is turnover net of inputs (leased inputs and/or interconnection).

**Recovery from end-users directly**

- most neutral economically is based on Ramsey mark-ups (to minimise distortions of demand and use) – impractical and inequitable.
- recovery via sales tax.
- recovery via contribution on value-added.

# Annex 2

# USO Net Cost Recovery - Illustration 2

Illustrative example of range of options for recovering a net USO cost (net of revenues and non-financial or indirect financial benefits) of (for example) 100 million euro per year.

## **Option A: Recovery from General Government budget**

Assumption: GDP 500 billion euro, total receipts of general government 40% of GDP (indirect taxes 11%, direct taxes 12%, social security contributions 13%, other 4%).

Net cost of USO would represent 0.05% of total general government receipts, (100 million euro out of total general government receipts of 200 billion euro).

## **Option B: Recovery from within Electronic Communications Sector**

### **Option B1: levy on undertakings directly**

Assumption: Turnover of telecommunications and related sectors of 20 billion euro. Excluding turnover of undertakings with less than 5 million euro annually, and also excluding double counted turnover (where undertakings compensate each other for leased infrastructure and for access and interconnection), gives assumed net turnover of 10 billion euro.

Net cost of USO would represent 1% of sector net turnover, (100 million euro out of net turnover of 10 billion euro).

### **Option B2: levy on final consumption**

Assumption: value-added (or final end user consumption) within sector of 10 billion euro (compared to total turnover of 20 billion euro). In cases where VAT rate is 20%, then basic levy on the sector would be 2 billion euro.

Net cost of USO would be equivalent to 1% of value-added within the sector, or an increment of 1 percentage point on the VAT levy, (100 million euro out of value-added sector of 10 billion euro)

Alternatively, 1 percentage point of the VAT levy could be retained for dealing with net cost of USO and the remainder transferred as a general government receipt. This would be equivalent to option A where the net USO cost effectively comes from general government tax receipts.